\documentclass[12pt,preprint]{aastex}
\shorttitle{SMA and $Spitzer$ Observations of CB\,17}%
\shortauthors{X. Chen et al.}%

\begin{document}

\title{SMA and $Spitzer$ Observations of Bok Glouble CB\,17: A Candidate First Hydrostatic Core?}

\author{Xuepeng~Chen\altaffilmark{1}, H\'{e}ctor~G.~Arce\altaffilmark{1}, Michael~M.~Dunham\altaffilmark{1},  Qizhou Zhang\altaffilmark{2},
Tyler~L.~Bourke\altaffilmark{2}, Ralf Launhardt\altaffilmark{3}, Markus Schmalzl\altaffilmark{3}, and Thomas Henning\altaffilmark{3}}

\affil{$^1$Department of Astronomy, Yale University, Box 208101, New Haven, CT 06520-8101, USA; xuepeng.chen@yale.edu}%
\affil{$^2$Harvard-Smithsonian Center for Astrophysics, 60 Garden Street, Cambridge, MA 02138, USA}%
\affil{$^3$Max Planck Institute for Astronomy, K\"{o}nigstuhl 17, D-69117 Heidelberg, Germany}%

\begin{abstract}

We present high angular resolution SMA and $Spitzer$ observations toward the Bok globule CB\,17. 
SMA 1.3\,mm dust continuum images reveal within CB\,17 two sources with an angular separation 
of $\sim$\,21$''$ ($\sim$\,5250\,AU at a distance of $\sim$\,250\,pc). The northwestern continuum 
source, referred to as CB\,17\,IRS, dominates the infrared emission in the $Spitzer$ images, drives 
a bipolar outflow extending in the northwest-southeast direction, and is classified as a low luminosity 
Class\,0/I transition object ($L_{\rm bol}$\,$\sim$\,0.5\,$L_\odot$). The southeastern continuum 
source, referred to as CB\,17\,MMS, has faint dust continuum emission in the SMA 1.3\,mm observations 
($\sim$\,6\,$\sigma$ detection; $\sim$\,3.8\,mJy), but is not detected in the deep $Spitzer$ infrared 
images at wavelengths from 3.6 to 70\,$\mu$m. Its bolometric luminosity and temperature, estimated 
from its spectral energy distribution, are $\leq$\,0.04\,$L_\odot$ and $\leq$\,16\,K, respectively. The SMA 
CO\,(2--1) observations suggest that CB\,17\,MMS may drive a low-velocity molecular outflow 
($\sim$\,2.5\,km\,s$^{-1}$), extending in the east-west direction. Comparisons with prestellar cores and 
Class\,0 protostars suggest that CB\,17\,MMS is more evolved than prestellar cores but less evolved than 
Class\,0 protostars. The observed characteristics of CB\,17\,MMS are consistent with the theoretical 
predictions from radiative/magneto hydrodynamical simulations of a first hydrostatic core, but there is
also the possibility that CB\,17\,MMS is an extremely low luminosity protostar deeply embedded in an
edge-on circumstellar disk. Further observations are needed to study the properties of CB\,17\,MMS 
and to address more precisely its evolutionary stage.

\end{abstract}


\section{INTRODUCTION}

Our knowledge of the formation and evolution of low-mass stars has made significant progress
over the past two decades (see, e.g., Reipurth et al.  2007 for several reviews). It is widely accepted 
that low-mass stars form from the gravitational collapse of dense molecular cores (e.g., Shu et 
al. 1987). Initially, these cores, generally referred to as prestellar cores, are cold dense condensations 
with infall motions, where no central stellar object yet exists (Andr\'{e} et al. 2009). Resulting from the 
collapse of prestellar cores, Class\,0 objects are the youngest accreting protostars observed right 
after point mass formation, when most of the mass of the system is still in the surrounding dense 
core/envelope (Andr\'{e} et al. 2000). Representing the earliest phase of low-mass star formation, 
both prestellar cores and Class\,0 protostars have been extensively observed and studied using 
(sub)\,millimeter and infrared telescopes (see, e.g., reviews by Di~Francesco et al. 2007; 
Ward-Thompson et al. 2007). However, despite all of the observational advances in the past two 
decades, we still do not have a good understanding of the evolutionary process that turns a 
prestellar core into a protostar. This is illustrated by the fact that several ``prestellar" cores, like L1014
and L1521F, were found to harbor very low-luminosity protostars in sensitive infrared observations (see
Young et al. 2004; Bourke et al. 2006). A better understanding of the star formation process can only 
be achieved by studying the detailed properties of cores and their surroundings at different evolutionary stages.

In this paper, we present high angular resolution observations of CB\,17, using the Submillimeter 
Array\footnote{The Submillimeter Array is a joint project between the Smithsonian Astrophysical 
Observatory and the Academia Sinica Institute of Astronomy and Astrophysics and is funded by the 
Smithsonian Institution and the Academia Sinica.} (SMA; Ho et al. 2004) and the {\it Spitzer Space 
Telescope} ($Spitzer$). CB\,17 (also known as L1389) is a small and slightly cometary-shaped dark 
cloud, located near Perseus and associated with the Lindblad ring (Lindblad et al. 1973). It was 
classified as a Bok globule by Clemens \& Barvainis (1988). The distance of CB\,17 is somewhat 
uncertain, ranging from $\sim$\,210\,pc (van Leeuwen 2007) to $\sim$\,300\,pc (Dame et al. 1987). 
Following Launhardt et al. (2010), we adopt a distance of 250\,$\pm$\,50\,pc in this work. 

The roundish core of the CB\,17 globule is associated with a faint and cold IRAS point 
source (IRAS\,04005+5647, detected only at 100\,$\mu$m and 60\,$\mu$m). CB\,17 
has been studied by various groups using different molecular line transitions (e.g., 
Lemme et al. 1996; Kane \& Clemens 1997; Launhardt et al. 1998; Benson et al. 1998; 
Turner et al. 1997, 2000; Caselli et al. 2002a). The core was found to have a mean kinetic 
gas temperature of $T_{\rm kin}\approx 10$\,K, and the observed non-thermal widths of 
optically thin lines are in the range 0.25--0.45\,km\,s$^{-1}$ (e.g., N$_2$H$^+$, Benson 
et al. 1998; Caselli et al. 2002a). Numerical simulations based on multi-line observations 
suggest that the kinematical structure of CB\,17 can be explained by a prestellar core 
with combined subsonic infall, small rotational, and low-level internal turbulent motions 
(Pavlyuchenkov et al. 2006). 

Based on the infrared and single-dish (sub-)\,millimeter continuum observations, Launhardt et 
al. (2010) found two sources within the CB\,17 globule. One source, referred to as CB\,17 IRS, 
dominates the infrared emission in the $Spitzer$ images but has faint millimeter continuum emission. 
The other source, referred to as CB\,17 SMM, has no compact infrared emission in the $Spitzer$ 
images but dominates the millimeter continuum emission. The results from fitting the spectral 
energy distributions (SEDs) suggested that CB\,17 IRS may be a Class\,0/I transition object while 
CB\,17 SMM may be a prestellar core. In addition, the IRAM-30m 1.3\,mm continuum images 
suggest that there may be two sub-cores (SMM\,1 and SMM\,2) in CB\,17 SMM with a separation 
of 14$''$, although this result was not confirmed by the less sensitive SCUBA 850\,$\mu$m images 
(see Launhardt et al. 2010).

\section{OBSERVATIONS AND DATA REDUCTION}

\subsection{SMA Observations}

The SMA 230\,GHz observations of CB\,17 were carried out in the compact configuration
on 2008 November 30 (eight antennas, $\sim$\,6.5 hours integration time) and 2009 
December 25 (seven antennas, $\sim$\,3.3 hours integration time). Zenith opacities during 
the observations were typically in the range of 0.10$-$0.15. In the 2008 observations, the 
digital correlator was set up to cover the frequency ranges 219.5$-$221.4 GHz and 
229.5$-$231.4 GHz in the lower and upper sidebands (LSB and USB), respectively. This 
setup includes the three isotopic CO lines of $^{12}$CO\,(2--1) (230.538\,GHz), 
$^{13}$CO\,(2--1) (220.399\,GHz), and C$^{18}$O\,(2--1) (219.560\,GHz), as well as 
N$_2$D$^+$\,(3--2) (231.322\,GHz). The channel widths were set up to be 0.406, 0.406,
0.203, and 0.203\,MHz for the four lines, corresponding to the velocity resolutions of
$\sim$\,0.50, $\sim$\,0.50, $\sim$\,0.25, and $\sim$\,0.25\,km\,s$^{-1}$, respectively.
The 1.3\,mm continuum emission was recorded with a total bandwidth of $\sim$\,3.5\,GHz, 
combining the line-free portions of the two sidebands ($\sim$\,1.7\,GHz USB and $\sim$\,1.8\,GHz LSB).
In the 2009 observations, the SMA bandwidth was upgraded to 4\,GHz, and the correlator 
was set up to cover the approximate frequency range of 216.8$-$220.8\,GHz in the LSB 
and of 228.8$-$232.8\,GHz in the USB. The total continuum bandwidth is $\sim$\,7.5\,GHz 
($\sim$\,3.8\,GHz LSB and $\sim$\,3.7\,GHz USB). System temperatures ranged from 100 
to 200\,K (depending on elevation) in the 2008 observations (typical value $\sim$\,140\,K) 
and from 100 to 150\,K in the 2009 observations (typical value $\sim$\,120\,K). The SMA 
primary beam is about 55$''$ at 230\,GHz.

The visibility data were calibrated with the MIR package (Qi 2005). In the 2008 observations, 
Saturn and quasar 3c454.3 were used for bandpass calibration, and quasars 0359+509 and 
3c111 for gain calibration. In the 2009 observations, quasar 3c273 was used for bandpass 
calibration, and quasars 3c84 and 0359+509 for gain calibration. Uranus and 3c273 were used 
for absolute flux calibration in the 2008 and 2009 observations, respectively. We estimate a 
flux accuracy of $\sim$\,20\%, by comparing the final quasar fluxes with the SMA calibration 
database. The calibrated visibility data were further imaged using the Miriad toolbox (Sault et 
al. 1995). We combined the 2008 and 2009 dust continuum data together to improve the 
imaging sensitivity. However, the line results shown in this paper were taken only from the 
2008 observations; the reason we did not include the 2009 line data is that the velocity 
resolution in the 2009 observations was set up to be about 1.0\,km\,s$^{-1}$ uniformly, about 
two times (in $^{12}$CO and $^{13}$CO) or four times (in C$^{18}$O and N$_2$D$^+$) larger 
than that of the 2008 line data. It must be noted that the line results taken in the 2008 and 2009 
observations are consistent with each other. Table~1 lists the SMA synthesized beam size and 
theoretical noise levels at 1.3\,mm continuum and in the CO\,(2--1) line, with robust {\it uv} weighting 1.0.

\subsection{$Spitzer$ Observations}

Infrared data of CB\,17 were obtained from the $Spitzer$ Science Center (SSC). CB\,17 was 
observed on 2004 February 12 with the Infrared Array Camera (IRAC; AOR key 4912384; 
PI: C. Lawrence) and 2004 October 16 with the Multiband Imaging Photometer for $Spitzer$ 
(MIPS; AOR key 12025088; PI: G. Rieke). The IRAC observations were taken in the high 
dynamic range mode, with an effective long-frame exposure time of $\sim$\,130\,s. The MIPS 
observations covered a field of $\sim$\,15$'$\,$\times$\,55$'$, with an exposure time of
$\sim$\,190\,s at 24\,$\mu$m, $\sim$\,80\,s at 70\,$\mu$m, and $\sim$\,20\,s at 160\,$\mu$m, 
respectively. The data were processed by the $Spitzer$ Science Center using their standard 
pipeline (version~S18.7 for IRAC data and version~S16.1 for MIPS data) to produce Post-Basic 
Calibration Data (P-BCD) images. The $Spitzer$ spatial resolution is about 2$''$ at the IRAC 
bands, and about 6$''$, 18$''$, and 40$''$ at the MIPS 24, 70, and 160\,$\mu$m bands, respectively. 
The overall flux calibration, expressed in physical units of MJy\,sr$^{-1}$, is estimated by the SSC 
to be accurate to within 10\%. Table~2 lists the imaging sensitivity (1\,$\sigma$) of the IRAC and 
MIPS bands for CB17. For comparison, the imaging sensitivity for dark cloud L1448 (c2d data; 
Evans et al. 2003; 2009) is also listed. The $Spitzer$ data of CB\,17 have been already published 
by Launhardt et al. (2010).

\section{RESULTS}

\subsection{Millimeter and Infrared Continuum Emission}

Figure~1 shows the SMA 1.3\,mm dust continuum image of CB\,17, in which a centrally-peaked 
continuum source is clearly detected in the northwest of the SMA field of view. This continuum 
source is spatially associated with the infrared source IRS found by Launhardt et al. (2010), which 
dominates the infrared emission in the $Spitzer$ images (see Figure~2) and is referred to as 
CB\,17\,IRS here. Located at the center of the SMA image, another faint continuum source 
is marginally detected (peak value $\sim$\,6\,$\sigma$; see Figure~1), at an angular distance of 
$\sim$\,21$''$ to the southeast of CB\,17\,IRS. This faint continuum source was independently 
detected in both the 2008 LSB ($\sim$\,5--6\,$\sigma$) and 2009 LSB ($\sim$\,4--5\,$\sigma$;
depending on the cleaning-box and $uv$-weighting adopted in the data reduction) observations,
but is not seen in the 2008 and 2009 USB observations. 
However, it must be noted that in our SMA observations the noises found in the USB are relatively 
higher than those in the LSB. This is evidenced by the observations of source CB\,17\,IRS, which was 
detected with the signal-to-noise ratios of $\sim$\,8 and $\sim$\,5 in the 2008 LSB 
(1\,$\sigma$\,$\sim$\,0.74\,mJy\,beam$^{-1}$) and 2009 LSB (1\,$\sigma$\,$\sim$\,0.65\,mJy\,beam$^{-1}$) 
observations, respectively, but only $\sim$\,5 and $\leq$\,3 in the 2008 USB (1\,$\sigma$\,$\sim$\,0.85\,mJy\,beam$^{-1}$) 
and 2009 USB (1\,$\sigma$\,$\sim$\,0.80\,mJy\,beam$^{-1}$) observations. Therefore, the faint continuum 
source found in the LSB observations is likely missing in the USB observations due to the relatively 
lower single-to-noise ratio. 

In the $Spitzer$ images shown in Figure~2, this faint dust continuum source is located at the center of 
a small-scale dark shadow seen in the IRAC 8.0\,$\mu$m image (see Figure~2a), and no compact infrared 
emission is detected from this source at the $Spitzer$ bands from 3.6 to 70\,$\mu$m (see Figure~2), 
although the MIPS 160\,$\mu$m map (spatial resolution $\sim$\,40$''$) shows a slight shift of the peak 
position from the northwestern IRS source toward this faint source. Hereafter, we refer to the central faint 
continuum source as CB\,17\,MMS. We also note that CB\,17\,MMS is spatially associated with a faint infrared 
source detected in the {\it Herschel} PACS 100\,$\mu$m observations\footnote{There is an offset ($\sim$\,2$''$) 
between the peaks of source CB\,17\,MMS and the faint infrared source seen at PACS 100\,$\mu$m (see
Figure~2b). This offset is smaller than the SMA synthesized beam size ($\sim$\,3$''$) and the PACS angular 
resolution at 100\,$\mu$m ($\sim$\,7$''$), and is therefore not significant.} (M.~Schmalzl et al., in preparation; 
see Figure~2b) and a core detected in various high-density molecular line tracers in CB\,17 (e.g., N$_{2}$H$^{+}$ 
and HCO$^{+}$; see Figure~2c). 
On the other hand, the IRAM-30m 1.3\,mm dust continuum observations show two sub-cores with $\sim$\,14$''$ 
separation in CB\,17, named SMM1 and SMM2 (see Launhardt et al. 2010 and also Figure~2d). Interestingly, 
in the high angular resolution SMA images, no continuum source is detected within the sub-core SMM2, while 
the SMA 1.3\,mm continuum source CB\,17\,MMS is close to the sub-core SMM1, but $\sim$\,7$''$ to the west 
of the peak position of the SMM1 core (see Figure~2d). This offset is larger than the pointing uncertainty in the 
IRAM-30m observations ($\sim$\,3$''$--5$''$). It is possible that the IRAM-30m single-dish observations detect 
the large-scale extended envelope, while the SMA observations reveal the relatively compact source which is 
embedded in this envelope, but not at its center.

Figure~3 shows the plots of the SMA visibility amplitudes versus {\it uv} distances for the continuum 
sources CB\,17\,IRS and MMS. As shown in the plots, CB\,17\,IRS shows a roughly flat distribution
of amplitudes from long to short baselines, suggesting a compact, point-like object. For CB\,17\,MMS, 
the distribution also shows a roughly flat distribution at baselines longer than $\sim$\,15\,k$\lambda$; 
at shorter baselines, the extended emission is mostly resolved (see below), and it is unclear if there is 
an Gaussian-like distribution, which is frequently seen in the interferometric observations toward 
protostellar envelopes (see, e.g., Looney et al. 2003). 
The SMA 1.3\,mm continuum fluxes of sources CB\,17\,IRS and MMS were derived from Gaussian fitting 
of the restored images using the Miriad command {\it imfit} (see Table~3). For comparison, their fluxes
integrated above the 3\,$\sigma$ level are also listed. For source CB\,17\,IRS, its flux detected at SMA is 
$\sim$\,6.3\,mJy, roughly 20\% of the flux found in the IRAM-30m image, where the flux density on its position 
is about $\sim$\,30\,mJy. For source CB\,17\,MMS,  its flux derived from the SMA images is $\sim$\,3.8\,mJy, 
while its flux detected in the IRAM-30m 1.3\,mm map is $\sim$\,50\,mJy (within the central 12$''$ or 3000\,AU 
of CB\,17\,MMS), indicating that more than 90\% of the flux around CB\,17\,MMS was resolved out by the SMA. 

Assuming that the 1.3\,mm dust continuum emission is optically thin, the total gas mass ($M_{\rm gas}$) 
is derived from the flux densities using the formula:

\begin{equation}
M_{\rm gas} = \frac{F_{\nu} D^{2}}
            {\kappa_{\rm m}(\nu)\, B_{\nu} (T_{\rm d})}\
\left(\frac{M_{\rm g}}{M_{\rm d}}\right) \quad
\end{equation}

\noindent where $D$\ is the distance to the source, $T_{\rm d}$ is the dust temperature, 
$\kappa_{\rm m}(\nu)$\ is the dust opacity (mass absorption coefficient per gram of dust), 
and $M_{\rm g}/M_{\rm d}$\ is the gas-to-dust mass ratio. Following Ossenkopf \& Henning 
(1994), we adopt $\kappa_{\rm m} = 0.5\,{\rm cm}^2\,{\rm g}^{-1}$, which is a typical value 
for dense cores with an average number density of $n(\rm H) = 10^5\,$cm$^{-3}$. A standard 
gas-to-dust mass ratio of 150 is used, which is a combination of the ratio of Hydrogen mass to 
dust mass of 110 (Draine \& Lee 1984) and the inclusion of Helium and heavier elements which 
introduces an extra factor of 1.36. A dust temperature of $\sim$\,10\,K was adopted for CB\,17\,MMS,
which is derived from a fit to the SED (see below and also Launhardt et al. 2010) and is also 
similar to the mean kinetic gas temperature found in CB\,17 (see $\S$\,1). For CB\,17\,IRS, a 
dust temperature of $\sim$\,20\,K is adopted (see below). The relative uncertainties of the derived 
masses due to the calibration errors of the fluxes are within $\pm$20\%. The total gas masses of 
CB\,17\,MMS and IRS derived from the SMA observations are $\sim$\,0.035 and $\sim$\,0.023\,$M_\odot$, 
respectively.

\subsection{CO\,(2--1) Emission}

Figure~4 shows the velocity channel maps of the SMA $^{12}$CO\,(2--1) emission. The $^{12}$CO 
emission is detected from $V_{\rm LSR}$\,=\,$-$7.2\,km\,s$^{-1}$ to $-$2.2\,km\,s$^{-1}$, with the 
cloud systematic velocity being $\sim$\,$-$4.7\,km\,s$^{-1}$ (Pavlyuchenkov et al. 2006). In each panel, 
the two crosses indicate the SMA positions of sources CB\,17\,IRS and MMS. In the CB\,17 core, the observed 
FWHM of the optically thin N$_{2}$H$^+$ line is in the range of 0.25--0.45\,km\,s$^{-1}$ (e.g., Benson et al. 
1998), and we can safely assume that the (turbulent) CO cloud emission is within a velocity width of 
$\sim$\,2\,km\,s$^{-1}$ ($\sim$\,5 times larger than the FWHM of N$_{2}$H$^+$). The $^{12}$CO emission 
at velocities more than 1\,km\,s$^{-1}$ away from the cloud systematic velocity suggests the existence of 
molecular outflows in this region (see $\S$\,4.2 for more discussions, and Arce et al. 2010 for a discussion 
on differentiating between outflows and turbulence-related features). For CB\,17\,IRS, blueshifted emission 
(from $V_{\rm LSR}$\,=\,$\sim$\,$-$7.2\,km\,s$^{-1}$ to $\sim$\,$-$5.7\,km\,s$^{-1}$) extends to the 
southeast, while redshifted emission (from $\sim$\,$-$3.7\,km\,s$^{-1}$ to $\sim$\,$-$2.7\,km\,s$^{-1}$)
mainly extends to the northwest of the source. Near CB\,17\,MMS, blueshifted emission 
(from $V_{\rm LSR}$\,=\,$\sim$\,$-$7.2\,km\,s$^{-1}$ to $\sim$\,$-$5.7\,km\,s$^{-1}$) extends to the east of 
the source, while redshifted emission (from $\sim$\,$-$3.7\,km\,s$^{-1}$ to $\sim$\,$-$2.2\,km\,s$^{-1}$) 
extends to both the west and east directions.

Figure~5 shows the velocity channel maps of the $^{13}$CO\,(2--1) emission of CB\,17, which is detected at 
velocities between $\sim$\,$-$5.2 and  $\sim$\,$-$4.2\,km\,s$^{-1}$. The $^{13}$CO\,(2--1) line emission
associated with CB\,17\,IRS shows a circular centrally-peaked condensation coincident with the dust continuum 
source, while the emission near CB\,17\,MMS extends in both the west and east directions, similar to the morphology 
of the $^{12}$CO\,(2--1) line emission. We also note that faint C$^{18}$O\,(2--1) emission is seen at the position of 
CB\,17\,IRS (see the spectrum in Figure~6), which shows a morphology similar to the $^{13}$CO\,(2--1) emission 
(see Figure~7), while no C$^{18}$O\,(2--1) emission is detected from source CB\,17\,MMS in our SMA observations.

\subsection{N$_2$D$^+$\,(3--2) Emission}

The N$_2$D$^+$\,(3--2) line emission is also detected in the CB\,17 observations. Figure~7 shows the 
velocity-integrated intensity map of the N$_2$D$^+$\,(3--2) emission, plotted on the IRAC 8.0\,$\mu$m 
image. The N$_2$D$^+$\,(3--2) emission shows an elongated structure, which extends roughly in the 
east-west direction, and is spatially coincident with the dark shadow seen in the IRAC 8.0\,$\mu$m image.
The FWHM diameter of the N$_2$D$^+$\,(3--2) condensation is measured to be 16.6$''$\,$\times$\,6.3$''$ or 
4200\,$\times$\,1600\,AU (at a distance of $\sim$\,250\,pc). Therefore, the condensation is resolved, 
as each axis is larger than the synthesized beam for N$_2$D$^+$ (3.0$''$\,$\times$\,2.8$''$). The HFS 
(HyperFine Structure) fitting routine in CLASS\footnote{see http://www.iram.fr/IRAMFR/GILDAS}, with
the frequencies and weights adopted from Dore et al. (2004), was used to derive LSR velocity ($V_{\rm LSR}$),
line width ($\Delta V$), optical depth ($\tau$), and excitation temperature ($T_{\rm ex}$). The optical depth 
($\tau$) is found to be small in most regions (ranging from 0.1 to 10, with typical values of 1--2). Hence, the 
N$_2$D$^+$ emission can be considered approximately optically thin, and the estimated excitation temperature 
is about 5--7\,K.
The measured integrated intensity of the N$_2$D$^+$\,(3--2) emission is about 5\,Jy\,beam$^{-1}$\,km\,s$^{-1}$
or $\sim$\,14\,K\,km\,s$^{-1}$. With the same method described in Caselli et al. (2002b), the column density of 
N$_2$D$^+$ is estimated to be $\sim$\,2--6\,$\times$\,10$^{11}$\,cm$^{-2}$.

The integrated intensity map of the C$^{18}$O\,(2--1) emission is also shown in Figure~7. Interestingly, the 
comparison between the N$_2$D$^+$\,(3--2), C$^{18}$O\,(2--1), and 8.0\,$\mu$m images suggests that 
N$_2$D$^+$\,(3--2) traces only cold gas around source CB\,17\,MMS, while C$^{18}$O\,(2--1) traces relatively
warm gas around infrared source CB\,17\,IRS.
Furthermore, no strong N$_2$D$^+$\,(3--2) emission is detected at the position of CB\,17\,MMS, but instead the 
N$_2$D$^+$\,(3--2) emission shows a small arc-like structure surrounding the MMS source. It could be that 
N$_2$D$^+$ at the CB\,17\,MMS position has been destroyed by the gradually warming gas/dust due to the 
accreting luminosity of source CB\,17\,MMS.

Figure~8 shows the velocity field of CB\,17, using the SMA N$_2$D$^+$\,(3--2) data and derived with the same 
method described in Chen et al. (2007). The mean velocity map shows a continuous velocity gradient across the 
N$_2$D$^+$ condensation, increasing from southeast to northwest (see Figure~8a). A least-squares fitting of 
the velocity field indicates there is a velocity gradient of 18$\pm$1\,km\,s$^{-1}$\,pc$^{-1}$, with a position angle 
of $\sim$\,$-$30$^\circ$ (measured east of north in the direction of increasing velocity), in the core traced by the
N$_2$D$^+$ emission.

The line widths are roughly constant within the condensation (with a typical value of $\sim$\,0.5\,km\,s$^{-1}$) and 
relatively large line widths (1.0--1.2\,km\,s$^{-1}$) are mainly seen at the southeast southeast edge (see Figure~8b).
We note that the velocity resolution in the N$_2$D$^+$ observations is $\sim$\,0.25\,km\,s$^{-1}$. Therefore, the 
typical line width of 0.5\,km\,s$^{-1}$ derived in the observations can be considered as an upper limit (given higher 
velocity resolution observations, we may expect to find narrower line widths between 0.25--0.50\,km\,s$^{-1}$).
Assuming a kinetic gas temperature of 10\,K (see $\S$\,1), the thermal contribution to the N$_2$D$^+$ line width is 
$\sim$\,0.13\,km\,s$^{-1}$, and the typical non-thermal contribution to the line width is then $\sim$\,0.48\,km\,s$^{-1}$, 
which is about 3.7 times larger than the thermal line width. Although the origin of the non-thermal line width is still a 
subject of an ongoing debate, it is widely accepted that turbulence is the main contributor (see, e.g., Goodman et al. 
1998). On the other hand, the thermal FWHM line width of an ``average" particle of mass 2.33\,m$_{\rm H}$ (assuming 
gas with 90\% H$_2$ and 10\% He), which represents the local sound speed, would be $\sim$\,0.44\,km\,s$^{-1}$ 
at 10\,K. The observed non-thermal line width in N$_2$D$^+$ is comparable with this value, which suggests that turbulence 
in the condensation is approximately sonic.

\section{DISCUSSION}

\subsection{Spectral Energy Distributions}

Figure~9 shows the spectral energy distribution of CB\,17\,IRS. The submillimeter and millimeter 
fluxes of CB\,17\,IRS were estimated from the SCUBA and IRAM-30m dust continuum images 
(see Launhardt et al. 2010). The fluxes in the $Spitzer$ images were measured with aperture 
photometry in the IRAF APPHOT package, with the radii, background annuli, and aperture 
corrections recommended by the $Spitzer$ Science Center. We note that CB\,17 is actually 
associated with a faint and cold IRAS point source 
(IRAS\,04005+5647, $F_{100\,\rm\mu m}=5.78$\,Jy, $F_{60\,\rm\mu m}=0.91$\,Jy, not detected 
at shorter wavelengths). Given that no compact infrared emission was detected from CB\,17\,MMS 
in the $Spitzer$ 3.6--70\,$\mu$m images but CB\,17\,IRS was detected at these bands, we assume 
that all the IRAS flux comes from CB\,17\,IRS\footnote{This is also consistent with the infrared observations 
taken with the {\it Herschel Space Observatory} (M.~Schmalzl et al. in preparation). In the $Herschel$ 
PACS observations at 100$\mu$m, source CB\,17\,IRS dominates the infrared emission (see Figure~2b),
with a flux ratio of more than 30 compared to source CB\,17\,MMS.}. 
Table~4 lists all the flux points of CB\,17\,IRS plotted in Figure~9. To derive luminosities and 
temperatures, we first interpolated and then integrated the SED, always assuming spherical 
symmetry. Interpolation between the flux densities was done by a $\chi$$^2$ single-temperature 
grey-body fit to all points (including upper limits) at $\lambda$\,$\geq$\,100\,$\mu$m, using 
the same method as described in Chen et al. (2008). A simple logarithmic interpolation was 
performed between all points at $\lambda$\,$<$\,100\,$\mu$m. The results from the SED 
fitting of CB\,17\,IRS, such as $T_{\rm bol}$\,$\sim$\,50\,K, $T_{\rm dust}$\,$\sim$\,20\,K, and 
$L_{\rm smm}$/$L_{\rm bol}$\,$\sim$\,2\%, in concert with the fact that it is directly observed 
in the near-infrared wavelengths (see Launhardt et al. 2010), suggest that CB\,17\,IRS is a 
Class\,0/I transition object with a luminosity of $L_{\rm bol}$\,$\sim$\,0.5\,$L_\odot$.

For comparison, we also show in Figure~9 the SED of source CB\,17\,MMS. The mm and submm 
fluxes of CB\,17\,MMS were estimated using the IRAM-30m and SCUBA dust continuum images 
(see Launhardt et al. 2010). The flux within one beam of CB\,17\,MMS is $\sim$\,50\,mJy at 1.3\,mm,
$\sim$\,180\,mJy at 850\,$\mu$m, and $<$\,800\,mJy at 450\,$\mu$m (3\,$\sigma$ upper limit, 
no detection). Although the MIPS\,3 image at 160\,$\mu$m does not resolve sources CB\,17\,IRS 
and MMS, a slight shift of the peak position from IRS toward MMS suggests detectable emission 
from MMS at this wavelength (see Figure~2d). For compiling the SED, we assigned 15\% of the 
total 160\,$\mu$m flux to MMS (see also Launhardt et al. 2010), but our results depend only weakly 
on the adopted flux splitting. Table~4 lists all the flux points and upper limits of CB\,17\,MMS. The 
estimated bolometric luminosity of CB\,17\,MMS is less than 0.04\,$L_\odot$, and the dust temperature 
derived from the SED fit is $\sim$\,10\,K. 

The fact that no infrared emission was detected from CB\,17\,MMS in the $Spitzer$ 3.6--70\,$\mu$m
images suggests this source is extremely cold; the estimated bolometric temperature of the source is 
about 16\,K. We note that the imaging sensitivity of the CB\,17 $Spitzer$ data is about 1.5 times deeper 
than that obtained by the c2d observations of other cores (see Table~2 for a comparison). Given that 
the c2d data are sensitive to embedded protostars with internal luminosity\footnote{The internal 
luminosity is the luminosity of the central source, which excludes the luminosity arising from external 
heating.} $L_{\rm int}$\,$\leq$\,4\,$\times$\,10$^{-3}$($d$/140\,pc)$^{2}$\,$L_\odot$ (see Dunham 
et al. 2008), the internal luminosity of a potential protostar in CB\,17\,MMS should be less than 0.013\,$L_\odot$ at 
a distance of 250\,pc, which is consistent with the upper limit of the bolometric luminosity obtained 
from our SED fitting ($\sim$\,0.04\,$L_\odot$). Nevertheless, it must be noted that uncertainties 
remain in our estimates, due to the limited observations available. More observations, such as high 
angular resolution and high sensitivity continuum observations at wavelengths from far-infrared to 
(sub-)\,millimeter, are needed to constrain the SED of CB\,17\,MMS in order to derive more precisely 
its luminosity and temperature.

\subsection{Outflows in CB\,17}

Figure~10 shows the velocity-integrated intensity map of the SMA $^{12}$CO\,(2--1) emission of CB\,17, 
plotted on the IRAC 8.0\,$\mu$m image. For source CB\,17\,IRS, its CO emission shows a typical bipolar 
morphology seen in low-mass protostellar outflows (see, e.g., Arce \& Sargent 2006; J{\o}rgensen et al. 
2007). From the SMA data, we estimate the opening angle of the blue lobe to be $\sim$\,85 degrees
and the position angle (measured east from north) is $\sim$\,125 degrees. The outflow's inclination (with 
respect to the plane of the sky) is estimated to be $\sim$\,50 degrees, using geometrical case~4 in Cabrit 
\& Bertout (1986).
Near source CB\,17\,MMS, the blueshifted and redshifted emissions show long and narrow structures 
($\sim$\,7500\,AU and $\sim$\,8500\,AU in length, respectively), extending in the east-west direction and 
overlapping each other, different from the morphology of the bipolar outflow associated with source
CB\,17\,IRS. Because CB\,17\,MMS is such an extremely low luminosity object, we discuss below five
possible mechanisms that might produce the observed CO emission around CB\,17\,MMS.

Firstly, the CO lobes around CB\,17\,MMS might be artifacts due to incomplete $uv$-coverage in the 
interferometric observations. However, the CO emission around CB\,17\,MMS is strong, and shows 
consistent velocity structure and morphology, which do not change significantly with data reduction 
methods (e.g., various $uv$ weighting, velocity resolution, and clean method adopted). More importantly, 
our observations taken in 2008 (eight antennas, 6.5 hours integration time) and 2009 (seven antennas, 
3.3 hours integration time) produce the same velocity structure and morphology around CB\,17\,MMS, 
indicating that these CO structures are independent of the $uv$ sampling. Therefore, we consider these 
CO lobes around CB\,17\,MMS are not artifacts in the interferometric maps.

Secondly, the low velocity ($\sim$\,2.5\,km\,s$^{-1}$) CO line emission might suggest that the emission 
could be due to bound motions in the cloud core. However, for gas at 2\,km\,s$^{-1}$ to be bound to the 
core at a distance of 7500\,AU (measured using the blue lobe), it would require a core mass of 
$\sim$\,17\,$M_\odot$, which is much larger than the core mass derived from the IRAM-30m continuum 
observations ($\sim$\,4\,$M_\odot$, radius 8000\,AU;  Launhardt et al. 2010) and the virial mass derived 
from molecular line observations ($\sim$\,3$M_\odot$, e.g., Caselli et al. 2002a; scaled to the radius of
7500\,AU). This indicates that the CO emission is not from spurious structures of the cloud core.

The third possibility is that the two long and narrow CO lobes around CB\,17\,MMS are caused by the outflow 
from CB\,17\,IRS, which impacts and deflects from the dense region near CB\,17\,MMS.
However, (1) the geometry of the system is not quite consistent with this picture. As seen in Figure~10,
the redshifted outflow of CB\,17\,IRS mainly extends to the northwest, but the red lobe around
CB\,17\,MMS mainly extends to the east. 
(2) The outflow from CB\,17\,IRS is much weaker than the elongated lobes close to CB\,17\,MMS (see 
Table~5). The velocity-integrared intensity of the CO emission from the two lobes around CB\,17\,MMS 
($\sim$\,19.5 and $\sim$\,23.3\,Jy\,beam$^{-1}$\,km\,s$^{-1}$ for the blue and red lobes, respectively) 
is much larger than that of the two lobes in the CB\,17\,IRS outflow 
($\sim$\,16.9 and $\sim$\,14.2\,Jy\,beam$^{-1}$\,km\,s$^{-1}$ for the blue and red lobes, respectively). 
We believe that it is unlikely for a weak outflow, like that of CB\,17\,IRS, to produce a much stronger 
outflow as a result from its deflection from a dense core.
And (3), if the CO emission around source CB\,17\,MMS is impacted and deflected from the CB\,17\,IRS
outflow, we might expect to find intense turbulence around CB\,17\,MMS, but in contrast, the dense core 
around CB\,17\,MMS is very quiescent, as indicated by the narrow linewidth in the optically thin N$_2$H$^+$ 
observations (see Bensen et al. 1998; Caselli et al. 2002a) and N$_2$D$^+$ observations (this work).

The fourth possibility is that the CO lobes around CB\,17\,MMS are actually parts of the outflow lobes of
CB\,17\,IRS, and these features could result from missing flux in the interferometric observations. However, 
the geometry of the system is not consistent with this picture. As we discussed above, the redshifted 
outflow of CB\,17\,IRS mainly extends to the northwest, but the red lobe around CB\,17\,MMS mainly 
extends to the east. Therefore, even if we assumed that the blue lobe around CB\,17\,MMS is part of the 
blueshifted outflow driven by CB\,17\,IRS, it would be difficult to explain the existence of the red lobe to the 
east of CB\,17\,MMS.

Finally, the fifth possibility is that the two CO lobes around source CB\,17\,MMS represent the molecular 
outflow driven by CB\,17\,MMS. The peculiar morphology (i.e., blue and red outflow lobes are overlapping) 
and the low radial velocity of the gas suggest that these blue- and redshifted lobes may be produced by a 
collimated outflow with an axis close to the plane of the sky, similar to the RNO\,43 outflow (e.g., Arce \&
Sargent 2005; see also Figure~1 of Cabrit et al. 1988 for a diagram of the geometry of an outflow close
to the plane of the sky).
If this is the case, then it is probable that we do not detect higher velocities CO emission due 
to projection effects (we return to this point in $\S$\,4.3.2). We consider this the most likely scenario, and 
hereafter we assume the observed narrow $^{12}$CO structures are associated with a molecular outflow 
driven by CB\,17\,MMS. Nevertheless, we note that further observations (e.g., SMA subcompact data) are 
needed to recover the missing flux of the extended structure, and to confirm the nature of these narrow 
$^{12}$CO structures.

Since the $^{13}$CO\,(2--1) emission from CB\,17\,MMS is detected only at velocities between $\sim$\,$-$5 
and $\sim$\,$-$4\,km\,s$^{-1}$ (see Figure~5), we may consider that the $^{12}$CO\,(2--1) emission at 
velocities beyond this range to be optically thin. The outflow masses of the two sources are derived with the 
same method as described in Cabrit \& Bertout (1990; 1992). In the calculations, we assume LTE conditions 
and an excitation temperature of 20\,K. The derived outflow mass, as well as other outflow properties (e.g.,
momentum $P$ and energy $E$), are listed in Table~5. The outflow mass-loss rate ($\dot{M}$$_{\rm out}$), 
force ($F_{\rm m}$), and mechanical luminosity ($L_{\rm m}$) are estimated from the mass, momentum, and 
energy with the estimated dynamical age of the outflow. For the CB\,17\,MMS outflow, assuming that the outflow 
velocity is 2.5\,km\,s$^{-1}$, the dynamical age of the outflow is estimated to be $\sim$\,1.4\,$\times$\,10$^4$ yr 
(the inclination effect is not considered here). For the CB\,17\,IRS outflow, assuming the same outflow velocity 
and an inclination angle of $\sim$\,50 degrees, the dynamical age of the outflow is estimated to be 
$\sim$\,1.1\,$\times$\,10$^4$ yr. We note that these outflow parameters in Table~5 refer only to the
compact outflows detected in the SMA maps and thus represent lower limits only.

\subsection{The Nature of CB\,17\,MMS}

\subsubsection{Comparisons to prestellar and Class\,0 objects}

Although the SMA and $Spitzer$ observations clearly indicate that source CB\,17\,IRS is a Class\,0/I 
transition object, the evolutionary stage of source CB\,17\,MMS is unclear. 
Previous single-dish molecular line observations have shown subsonic infall ($\sim$\,0.05\,km\,s$^{-1}$), 
slow rotation ($\sim$\,2\,km\,s$^{-1}$\,pc$^{-1}$), and subsonic internal turbulent ($\sim$\,0.1\,km\,s$^{-1}$) 
motions in the CB\,17 dense core (see Pavlyuchenkov et al. 2006), which are similar to the typical properties 
found in prestellar cores (see, e.g., Andr\'{e} et al. 2009). These early results of CB\,17\footnote{The previous 
single-dish line observations toward CB\,17 generally show a dense core centered at the position of source 
CB\,17\,MMS. Therefore, we consider that the properties derived from these single-dish line observations are 
related to source MMS rather than source IRS.}, in concert with the fact that no compact infrared emission is 
detected from CB\,17\,MMS in the $Spitzer$ images, led to the idea that CB\,17\,MMS may be a prestellar core 
(e.g., Launhardt et al. 2010). 
However, the SMA 1.3\,mm dust continuum observations suggest that a compact object has formed in 
CB\,17\,MMS. In contrast, prestellar cores generally show no compact dust continuum emission in high 
angular resolution interferometric observations (e.g., Olmi et al. 2005; Schnee et al. 2010; but see also
Bourke et al. 2012). Furthermore, the SMA CO\,(2--1) observations suggest that CB\,17\,MMS drives a 
molecular outflow, which implies active accretion/ejection motions in source CB\,17\,MMS --- unlikely to 
take place in a prestellar core.
Moreover, Pavlyuchenkov et al. (2006) suggested that the CB\,17 core is chemically evolved, with an age 
of $\sim$\,2\,Myr, which is somewhat larger than the typical lifetimes of prestellar cores 
($\sim$\,1--1.5\,$\times$\,10$^6$\,yr; see Andr\'{e} et al. 2009).

On the other hand, compared to the typical properties of Class\,0 protostars, CB\,17\,MMS shows also
three major differences. Firstly, CB\,17\,MMS is not visible in the deep $Spitzer$ images at wavelengths 
from 3.6 to 70\,$\mu$m, with an extremely low bolometric luminosity ($\leq$\,0.04\,$L_\odot$), which 
implies no central protostellar object formed yet within the dense core. For comparison, most Class\,0 
protostars, if not all, are detectable in the $Spitzer$ infrared images (at least in the MIPS bands; see, 
e.g., Evans et al. 2009). Secondly, the SMA 1.3\,mm dust continuum observations suggest the existence 
of a compact, but very faint, object in CB\,17\,MMS (see $\S$\,3.1). The mm continuum flux of CB\,17\,MMS 
(only a few mJy) is about two magnitudes lower than the values of those Class\,0 protostars observed 
with the same configuration at the SMA (generally in the order of $\sim$\,100\,mJy or $\sim$\,0.1\,$M_\odot$;
see, e.g., J{\o}rgensen et al. 2007), implying that no massive accretion disk has developed yet around 
CB\,17\,MMS. Lastly, CB\,17\,MMS appears to drive a relatively slow molecular outflow 
($\sim$\,2.5\,km\,s$^{-1}$), compared to those typically found in Class\,0 protostars, which have velocities of 
$\sim$\,10 to 100\,km\,s$^{-1}$ and have strong  (physical and chemical) impacts on their surrounding cores 
and clouds (see, e.g., Arce et al. 2007).

From the comparisons discussed above, we suggest that source CB\,17\,MMS is more evolved than prestellar 
cores but less evolved than Class\,0 protostars, as it preserves many typical properties of prestellar cores seen 
in the early single-dish observations (i.e., in the outer core), but it also shows properties (e.g., compact object 
and outflow) that only protostars exhibit, in the high angular resolution observations (i.e., in the inner core), 
though neither strong protostellar object nor massive accretion disk appear to have developed yet in it.

\subsubsection{A Candidate First Hydrostatic Core?}

Interestingly, the observed properties of CB\,17\,MMS are consistent with the theoretical predictions of the radiative 
hydrodynamical (RHD) simulations for the first hydrostatic core (or first core), a transient object intermediate between 
the prestellar and Class\,0 phases (see, e.g., Larson 1969; Masunaga et al. 1998). Theoretical studies have investigated 
the properties of the first core and made a series of observationally testable predictions, including a short lifetime of 
only 10$^3$$-$10$^4$ years, an extremely low bolometric luminosity ($<$\,0.1\,$L_\odot$), very low mass ($<$\,0.1\,$M_\odot$), 
and no detectable infrared emission at wavelengths shorter than 30\,$\mu$m (see, e.g., Boss \& Yorke 1995; Masunaga 
et al. 1998; Machida et al. 2008).
In Figure~11, we compare the fluxes of source CB\,17\,MMS with those of a first hydrostatic core modeled by 
Masunaga et al. (1998). Although the number of observational data points is still small, the comparisons show 
that the fluxes of CB\,17\,MMS are in general agreement with those of a first hydrostatic core calculated with a 
beam size of 1000\,AU (or 4$''$ at the distance of CB\,17), which would have a luminosity of $\sim$\,0.06\,$L_\odot$ 
when the core central density reaches $\rho$$_{\rm center}$ $\sim$ 10$^{-9}$\,g\,cm$^{-3}$ or evolves at 
$\sim$\,1.23\,$\times$\,free-fall time (model M1a; see Masunaga et al. 1998 for more details).

In addition, the radius of the first core is found to be small, typically $\sim$\,5\,AU (Masunaga et al. 1998). 
When the effects of rotation are considered, this radius can be a little larger, in the range of 10$-$20\,AU 
(Saigo \& Tomisaka 2006; Saigo et al. 2008). Generally, in the interferometric dust continuum observations, 
prestellar cores show no compact dust continuum emission (i.e. no detection) with existing millimeter interferometers 
(e.g., Olmi et al. 2005; Schnee et al. 2010; Offner et al. 2012; X.~Chen et al. in preparation), 
because the extended structure of prestellar core (typically with density of $\rho$\,$\sim$\,10$^{-19}$\,g\,cm$^{-3}$, 
or $\sim$\,10 magnitudes smaller than that of the first core) will be almost totally resolved out by millimeter 
interferometric observations. While a core evolves from prestellar core to the onset of the first core, we expect to 
find a relatively compact, point-like, but also faint ($\leq$\,50\,mJy, judged from the models in Masunaga et al. 
1998) object in the interferometer observations. As can be seen in Figure~3, the SMA visibility amplitudes versus 
{\it uv} distances diagram of CB\,17\,MMS shows a roughly flat distribution, which suggests a point-like object 
in CB\,17\,MMS with a flux of $\sim$\,3\,mJy, consistent with the prediction of the small radius and small flux of the first 
core. However, we note that the SMA compact configuration observations mainly sample the $uv$ range between 
$\sim$\,10--60\,k$\lambda$. Further subcompact and extended configurations are definitely needed to study in 
detail the density structure of CB\,17\,MMS.

Moreover, the SMA CO\,(2--1) observations suggest that CB\,17\,MMS drives a molecular outflow. 
Interestingly, recent MHD simulations have shown that the first cores can drive low-velocity outflows (see, e.g., 
Machida et al. 2008; Tomida et al. 2010). In the simulations, the typical outflow driven by a first core has velocities 
of $\sim$\,2--5\,km\,s$^{-1}$ (e.g., Machida et al. 2008). The observed velocity of the CB\,17\,MMS outflow is 
about 2.5\,km\,s$^{-1}$, which is in good agreement with the result from the MHD simulations. However, in the MHD 
simulations (e.g., Machida et al. 2008), the outflow driven by the first core also features wide opening-angle and 
low extent-to-width ($E$/$W$) ratio (2.2$-$2.5)\footnote{More recently, smoothed particle magneto-hydrodynamics 
simulations suggest that the first hydrostatic core can drive collimated jets (opening angles $\leq$\,10$^\circ$) with 
speeds of $\sim$\,2--7\,km\,s$^{-1}$ (see Price et al. 2012).}. 
The outflow driven by CB\,17\,MMS shows two narrow lobes, overlapping with each other, with opening angle of 
$\sim$\,35$^\circ$ and $E$/$W$ of $\sim$\,4 (measured using the blue lobe), which is more consistent with a 
protostellar outflow rather than a first core outflow. However, it must be noted that the opening-angle and $E$/$W$ 
measured in the CB\,17\,MMS outflow are best treated as upper limits, because the SMA compact configuration 
observations recover only the fluxes at projected baselines $>$\,10\,k$\lambda$ (corresponding to angular scales 
$<$\,20$''$). Further short-spacing observations are needed to recover the extended structure of the CB\,17\,MMS 
outflow, in order to derive more precisely its opening-angle and $E$/$W$ ratio. In MHD simulations, both magnetic 
field and rotation rate in the collapsing core shape the morphology of the first core outflow (see, e.g., Machida et al. 
2008). Hence, there is also the possibility that the CB\,17\,MMS outflow represent a first core outflow that results from 
a specific magnetic field and/or rotation rate.

Another significant unknown factor in our analysis of the CB\,17\,MMS outflow is the source inclination. With the SMA 
CO images of CB\,17\,MMS, we are unable to set strong constraints on the inclination from the outflow morphology, 
although the extreme face-on ($\leq$\,10$^\circ$) configuration can be ruled out here. If CB\,17\,MMS is viewed 
along a relatively edge-on line-of-sight ($>$\,60$^\circ$), the true velocity of the outflow would increase beyond the 
outflow velocity expected for the first core. For example, an inclination of 80$^\circ$ would increase the outflow velocity 
from the observed value of 2.5 to 14\,km\,s$^{-1}$. On the other hand, if we assumed an inclination of 45$^\circ$, the 
outflow velocity of CB\,17\,MMS would be about 3.5\,km\,s$^{-1}$, which is still comparable with the predicted values of 
the first core outflows (but in this case, the extent-to-width ratio would increase to $\sim$\,6).
Furthermore, source inclination also has strong effects on the observed infrared properties. 
For a protostar surrounded by a circumstellar disk and embedded in a dense envelope, the 
infrared emission would be much stronger when viewed near face-on where the emission can 
escape through outflow cavities, than when viewed near edge-on where the emission is 
reprocessed by the disk and dense inner envelope. Clearly, better knowledge of the source 
inclination would help in determining the true evolutionary status of CB\,17\,MMS.

At present, unfortunately, due to the insufficient observations and uncertain inclination angle, the nature of source 
CB\,17\,MMS is still not definitive. With the combined results from early single-dish observations and our SMA and 
$Spitzer$ observations, in concert with the comparisons to theoretical models, we consider that CB\,17\,MMS may 
represent a candidate first hydrostatic core. Nevertheless, there is also the possibility that CB\,17\,MMS is an extremely 
low luminosity protostar, which is deeply embedded in an {\it edge-on} circumstellar disk/inner envelope and thus shows 
no detectable infrared emission in the deep $Spitzer$ observations.

\subsubsection{Comparisons with other first core candidates}

The detection of the first hydrostatic core phase is of prime importance in our understanding of the early evolution of 
dense cores and the origin of outflows, as it would not only confirm the long prediction of RHD models but also set 
strong constraints on MHD models of protostellar outflows. 
On the observational side, the search for the first core has been going on for a while and several first core candidates 
have been proposed, although none of them have been verified.
Based on the SMA and $Spitzer$ observations, we reported a first core candidate, L1448~IRS\,2E, which is 
invisible in the sensitive $Spitzer$ infrared images (from 3.6 to 70\,$\mu$m), has very weak (sub-)\,millimeter dust 
continuum emission, and consequently has an extremely low luminosity ($L_{\rm bol}$ $<$\,0.1\,$L_\odot$), but 
also drives a molecular outflow (Chen et al. 2010).
Enoch et al. (2010) reported another candidate first core, Per-Bolo~58, a very low luminosity (internal luminosity 
$<$\,0.01\,$L_\odot$) dense core in Perseus. This core was originally thought to be starless, but Enoch et al. (2010) 
detected an associated infrared source in very deep $Spitzer$ 24\,$\mu$m and 70\,$\mu$m observations and argued 
this source could either be a first core or a very low luminosity Class\,0 protostar. A bipolar, jet-like molecular 
outflow was also found in this interesting source by Dunham et al. (2011). 
More recently, Pineda et al. (2011) reported another candidate, L1451-mm, which is not visible in the $Spitzer$ 
observations (with a bolometric luminosity of $<$\,0.05\,$L_\odot$) and drives a low-velocity, poorly-collimated, 
bipolar outflow.

Together with CB\,17\,MMS, these sources are all suggested candidates of the long predicted first hydrostatic 
core, but none of them are in complete agreement with theoretical models. For source L1448 IRS\,2E, its outflow 
velocity reaches $\sim$\,25\,km\,s$^{-1}$ (see Chen et al. 2010), much larger than the value predicted by the 
MHD simulations. In the case of Per-Bolo~58, its detection at 24\,$\mu$m is inconsistent with current models. 
For source L1451-mm, its observations are in better agreement with theoretical models, but its SED 
and continuum interferometric visibilities can also be equally well fitted by a model of a protostar plus a circumstellar
disk (see Pineda et al. 2011). For source CB\,17\,MMS, another promising candidate, its outflow is much more 
collimated than that expected for a first core, and there is a possibility that it is an extremely low luminosity protostar 
deeply embedded in an edge-on circumstellar disk.

Clearly, more observations are needed to study these sources thoroughly: (1) high angular resolution and high 
sensitivity continuum observations at wavelengths from far-infrared to (sub-)\,millimeter, are needed to constrain 
the SEDs of these sources, in order to derive more precisely their luminosities and temperatures, and (2) high 
angular and spectral resolution line observations are also needed to study in detail the physical properties, density 
structure, kinematics, and chemistry of the surrounding dense gas around the sources in order to accurately 
characterize their evolutionary status. It is also critical to search for more candidate first cores in nearby clouds, 
in order to achieve a better understanding of the formation and evolution of dense cores, as well as the origin of 
outflows. With the availability of recent sensitive (sub-)\,millimeter telescopes (e.g., {\it Herschel Space Observatory} 
and the Atacama Large Millimeter/Submillimeter Array), we believe that more candidates will be found in the near future.

\section{SUMMARY}

We present SMA 230\,GHz and $Spitzer$ infrared observations toward the Bok globule CB\,17.
The SMA 1.3\,mm dust continuum images reveal within CB\,17 two sources, which are separated 
by $\sim$\,21$''$ ($\sim$\,5250\,AU) in the northwest-southeast direction. The northwestern continuum 
source, referred to as CB\,17\,IRS (gas mass $\sim$\,0.023\,$M_\odot$), dominates the infrared emission 
in the $Spitzer$ images and drives a low-velocity bipolar outflow detected in the SMA CO\,(2--1) observations. 
The SED fitting results suggest that CB\,17\,IRS is a low luminosity Class\,0/I transition object 
($L_{\rm bol}$\,$\sim$\,0.5\,$L_\odot$). 

The southeastern continuum source, referred to as CB\,17\,MMS, has faint dust continuum emission in 
the SMA 1.3\,mm images ($\sim$\,6\,$\sigma$ detection; gas mass $\sim$\,0.035\,$M_\odot$), and is not 
detected in the deep $Spitzer$ images at wavelengths from 3.6\,$\mu$m to 70\,$\mu$m. Its bolometric 
luminosity and temperature, estimated from the SED fitting, are $\leq$\,0.04\,$L_\odot$ and $\leq$\,16\,K, 
respectively. The SMA N$_2$D$^+$\,(3--2) observations show an elongated condensation associated
with source CB\,17\,MMS, which has a systematic velocity gradient of 18$\pm$1\,km\,s$^{-1}$\,pc$^{-1}$
and a typical line width of $\sim$\,0.48\,km\,s$^{-1}$.
Interestingly, the SMA CO\,(2--1) observations suggest that CB\,17\,MMS may drive a long
narrow low-velocity outflow ($\sim$\,2.5\,km\,s$^{-1}$), with blueshifted and redshifted lobes extending in 
the east-west direction, that overlap each other. Comparisons with prestellar cores and Class\,0 
protostars suggest that CB\,17\,MMS is likely at an evolutionary stage intermediate between these two
stages. The observed characteristics of CB\,17\,MMS are consistent with the properties expected for the 
first hydrostatic core as predicted by radiative/magneto hydrodynamical simulations. We thus consider 
CB\,17\,MMS to be a candidate first core. However, there is also the possibility that CB\,17\,MMS is an
extremely low luminosity protostar, which is deeply embedded in an edge-on circumstellar disk/inner
envelope. Further high angular resolution and high sensitivity observations are needed to confirm the 
properties of CB\,17\,MMS and to address more precisely its evolutionary stage.

\acknowledgments
We thank the SMA staff for technical support during the observations and 
the $Spitzer$ Science Center for their maintenance of the $Spitzer$ data.
This material is based on work supported by NSF grant AST-0845619 to 
H.G.A. This research is supported in part by the National Science Foundation 
under grant number 0708158 (T.L.B.).

\clearpage

\clearpage


\begin{deluxetable}{lccccc}
\tabletypesize{\scriptsize}%
\tablecaption{\footnotesize SMA Observations of CB\,17%
\label{sma_log}}%
\tablewidth{0pt}%
\tablehead{\colhead{Object} &\colhead{R.A. \& Dec. (J2000)$^{a}$}
&\colhead{Distance}&\colhead{Array}
&\colhead{HPBW$^{b}$} &\colhead{rms$^c$}\\
\colhead{Name}&\colhead{[h\,:\,m\,:\,s,
$^{\circ}:\,':\,''$]}&\colhead{[pc]}&\colhead{configuration}
&\colhead{[arcsec]} &\colhead{[mJy\,beam$^{-1}$]}}\startdata

CB\,17      &  04:04:35.85, 56:56:03.09  &  250 &  compact  &  3.1$\times$2.8  & 0.55, 55\,(0.5\,km\,s$^{-1}$)\\

\enddata
\tablenotetext{a}{Phase center in the SMA observations.}%
\tablenotetext{b}{Robust (+1.0) weighted synthesized FWHM beam size of the SMA 1.3\,mm dust continuum map (combining 
the 2008 and 2009 LSB data).}%
\tablenotetext{c}{1\,$\sigma$ theoretical noises of 1.3\,mm dust continuum and the $^{12}$CO\,(2--1) line (per channel width).}
\end{deluxetable}

\begin{deluxetable}{lccccccc}
\tabletypesize{\scriptsize} \tablecaption{\footnotesize $Spitzer$
Imaging Sensitivities of CB\,17 \& L1448$^a$\label{spitzer}}
\tablewidth{0pt} \tablehead{\colhead{Source} &\colhead{IRAC1}
&\colhead{IRAC2} &\colhead{IRAC3} &\colhead{IRAC4}
&\colhead{MIPS1}
&\colhead{MIPS2}&\colhead{MIPS3}\\
\colhead{} &\colhead{(3.6\,$\mu$m)} &\colhead{(4.5\,$\mu$m)}
&\colhead{(5.8\,$\mu$m)} &\colhead{(8.0\,$\mu$m)}
&\colhead{(24\,$\mu$m)}
&\colhead{(70\,$\mu$m)}&\colhead{(160\,$\mu$m)}\\
\hline \colhead{} &\colhead{[MJy/sr]} &\colhead{[MJy/sr]}
&\colhead{[MJy/sr]} &\colhead{[MJy/sr]}
&\colhead{[MJy/sr]}&\colhead{[MJy/sr]}&\colhead{[MJy/sr]}}\startdata
CB\,17   & 0.10$\pm$0.03 & 0.17$\pm$0.04  & 0.8$\pm$0.1   & 3.0$\pm$0.2   & 22$\pm$\,0.2  & 14$\pm$2 & 100$\pm$10\\
L1448    & 0.20$\pm$0.05 & 0.25$\pm$0.05  & 1.5$\pm$0.2   & 6.5$\pm$0.5   & 36$\pm$\,1.0  & 20$\pm$5 & 150$\pm$20\\
\hline
         & [$\mu$Jy/pixel]&[$\mu$Jy/pixel]&[$\mu$Jy/pixel]&[$\mu$Jy/pixel]&[mJy/pixel]& [mJy/pixel]&[mJy/pixel]\\
\hline
CB\,17   & 3.6$\pm$1.0  & 6.1$\pm$1.4  & 29$\pm$3   & 110$\pm$7    & 3.6$\pm$0.03       & 7$\pm$1  & 600$\pm$60\\
L1448    & 7.2$\pm$2.0  & 9.0$\pm$1.8  & 54$\pm$6   & 240$\pm$20   & 5.9$\pm$0.15       & 10$\pm$3 & 900$\pm$90\\
\enddata
\tablenotetext{a}{1\,$\sigma$ rms of images near the positions of the two sources.}
\end{deluxetable}

\begin{deluxetable}{lcccccc}
\tabletypesize{\scriptsize} \tablecaption{\footnotesize SMA
1.3\,mm dust continuum results of CB\,17\label{sma dust continuum}} \tablewidth{0pt} \tablehead{\colhead{Source}
&\colhead{R.A. \& Dec. (J2000)} &
\multicolumn{2}{c}{Gaussian Fitting}&\colhead{}& \multicolumn{2}{c}{3\,$\sigma$ Level$^b$}\\
\cline{3-4}\cline{6-7}\colhead{Name}   & \colhead{[h\,:\,m\,:\,s,
$^{\circ}:\,':\,''$]} & \colhead{Flux (mJy)}& \colhead{$M_{\rm
gas}$$^a$ ($M_\odot$)}&\colhead{} & \colhead{Flux (mJy)}&
\colhead{$M_{\rm gas}$$^a$ ($M_\odot$)}} \startdata

CB\,17\,IRS     &  04:04:33.76, 56:56:16.5 & 6.3$\pm$1.3 & 0.023$\pm$0.005 && 5.8$\pm$1.2 & 0.021$\pm$0.004 \\

CB\,17\,MMS   & 04:04:35.78, 56:56:03.4  & 3.8$\pm$0.7  & 0.035$\pm$0.007 && 3.5$\pm$0.7 & 0.032$\pm$0.006\\

\enddata
\tablenotetext{a}{Total gas mass; See text for dust temperature and opacity used.}
\tablenotetext{b}{Flux integrated above the 3\,$\sigma$ level, and corresponding gas mass.}
\end{deluxetable}

\begin{deluxetable}{lccccc}
\tabletypesize{\scriptsize} \tablecaption{\footnotesize Photometry of CB\,17\,IRS and MMS\label{photometry}} \tablewidth{0pt}%
\tablehead{\colhead{}&\multicolumn{2}{c}{CB\,17\,IRS}&\colhead{}&\multicolumn{2}{c}{CB\,17\,MMS}\\
\cline{2-3}\cline{5-6}\colhead{$\lambda$}&\colhead{$S_{\nu}$}&\colhead{Aperture} &\colhead{} &\colhead{$S_{\nu}$}&\colhead{Aperture}\\
\colhead{($\mu$m)}&\colhead{(mJy)}&\colhead{(arcsec)} &\colhead{} &\colhead{(mJy)}&\colhead{(arcsec)}}\startdata

2.2   & 1.1$\pm$0.5     & 5.0 & & \nodata        & \nodata   \\
3.6   & 1.71$\pm$0.3   & 2.4 & & $<$\,0.011$^a$   & 1.2   \\
4.5   & 1.78$\pm$0.3   & 2.4 & & $<$\,0.018$^a$   & 1.2   \\
5.8   & 1.22$\pm$0.3   & 2.4 & & $<$\,0.084$^a$   & 1.2   \\
8.0   & 0.6$\pm$0.2     & 2.4 & & $<$\,0.3$^a$       & 1.2   \\
24    & 92$\pm$20       & 5    & & $<$\,11$^a$        & 2.5  \\
60    & 910$\pm$450   & 75  & & \nodata                & \nodata   \\
70    & 1130$\pm$220   & 20& & $<$\,90$^a$        & 20   \\
100  & 5780$\pm$2000  & 125 & & 36$\pm$2$^b$ & 7  \\
160   & 4300$\pm$1500   & 40 & & 800$\pm$200   & 20   \\
450   & $<$\,1000$^c$   & 10 & & $<$\,800$^c$     & 10   \\
850   & $<$\,120$^c$   & 14  & & 180$\pm$40$^c$   & 14  \\
1300  & 30$\pm$6$^c$   & 12  & & 50$\pm$10$^c$    & 12  \\

\enddata
\tablenotetext{a}{Detection upper limits (3\,$\sigma$ per aperture) at the position of CB\,17\,MMS in the $Spitzer$ images.}%
\tablenotetext{b}{Flux derived from the $Herschel$ PACS 100\,$\mu$m observations (M.~Schmalzl et al. in preparation).}%
\tablenotetext{c}{Flux estimated using dust continuum emission within one beam of CB\,17\,IRS
and MMS in the JCMT/SCUBA and IRAM-30m images.} %
\end{deluxetable}

\begin{deluxetable}{lcccccc}
\tabletypesize{\scriptsize}\tablecaption{\footnotesize Outflow
parameters of CB\,17\,IRS and MMS$^{a}$\label{outflow}}
\tablewidth{0pt}
\tablehead{\colhead{}&\colhead{Mass}&\colhead{Momentum}&\colhead{Energy}
&\colhead{$\dot{M}$$_{\rm out}$}&\colhead{Force}&\colhead{Luminosity}\\
\colhead{Component}&\colhead{[10$^{-3}$\,$M_\odot$]}&\colhead{[10$^{-3}$$M_\odot$\,km\,s$^{-1}$]}
&\colhead{[10$^{40}$\,ergs]}&\colhead{[10$^{-8}$\,$M_\odot$\,yr$^{-1}$]}
&\colhead{[10$^{-8}$\,$M_\odot$\,km\,s$^{-1}$\,yr$^{-1}$]}&\colhead{[10$^{-5}$$L_\odot$]}}
\startdata

CB\,17\,IRS blue   & 0.30 & 0.75  & 1.9 & 2.7 & 6.8 & 1.4\\
CB\,17\,IRS red    & 0.22 & 0.49  & 1.1 & 2.0 & 4.5 & 0.8\\
CB\,17\,MMS blue & 0.33 & 0.84  & 2.1 & 2.4 & 6.0 & 1.2\\
CB\,17\,MMS red  & 0.46 & 1.02  & 2.2 & 3.3 & 7.3 & 1.3\\

\enddata
\tablenotetext{a}{~Lower limits derived from the SMA CO\,(2--1) observations.} 
\end{deluxetable}

\clearpage

\begin{figure*}
\begin{center}
\includegraphics[width=10cm,angle=0]{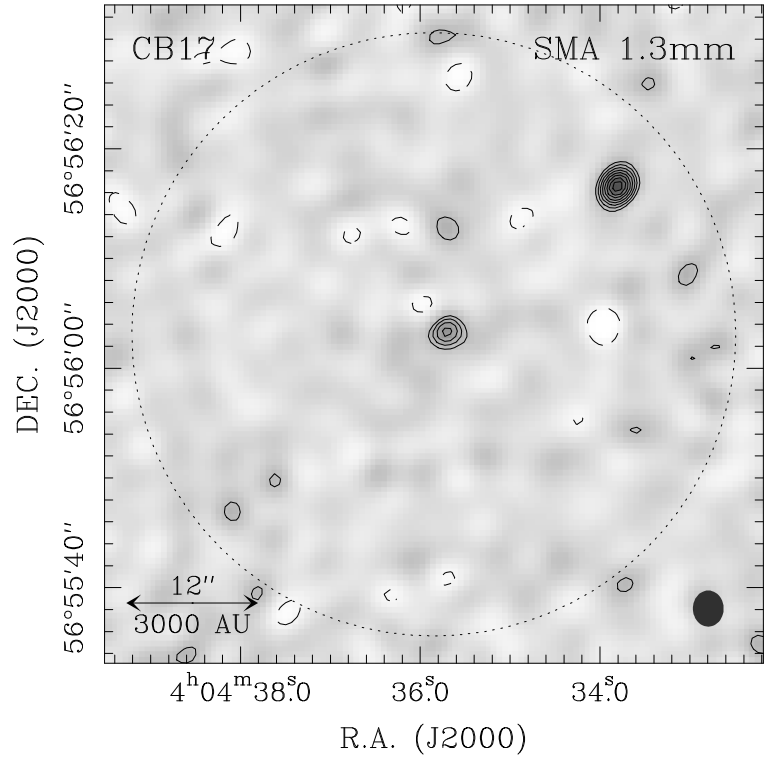}
\caption{The SMA 1.3\,mm dust continuum image of CB\,17, using the data combining the 2008 and 2009 LSB observations. 
The contours start at $\pm$3\,$\sigma$, and then increase from +3\,$\sigma$ in steps of 1\,$\sigma$ ($\sim$\,0.55\,mJy\,beam$^{-1}$). 
The synthesized SMA beam is shown as a grey oval in the bottom right corner. The dotted circle shows the SMA primary beam size ($\sim$\,55$''$).
\label{cb17_sma}}
\end{center}
\end{figure*}


\begin{figure*}
\begin{center}
\includegraphics[width=14cm,angle=0]{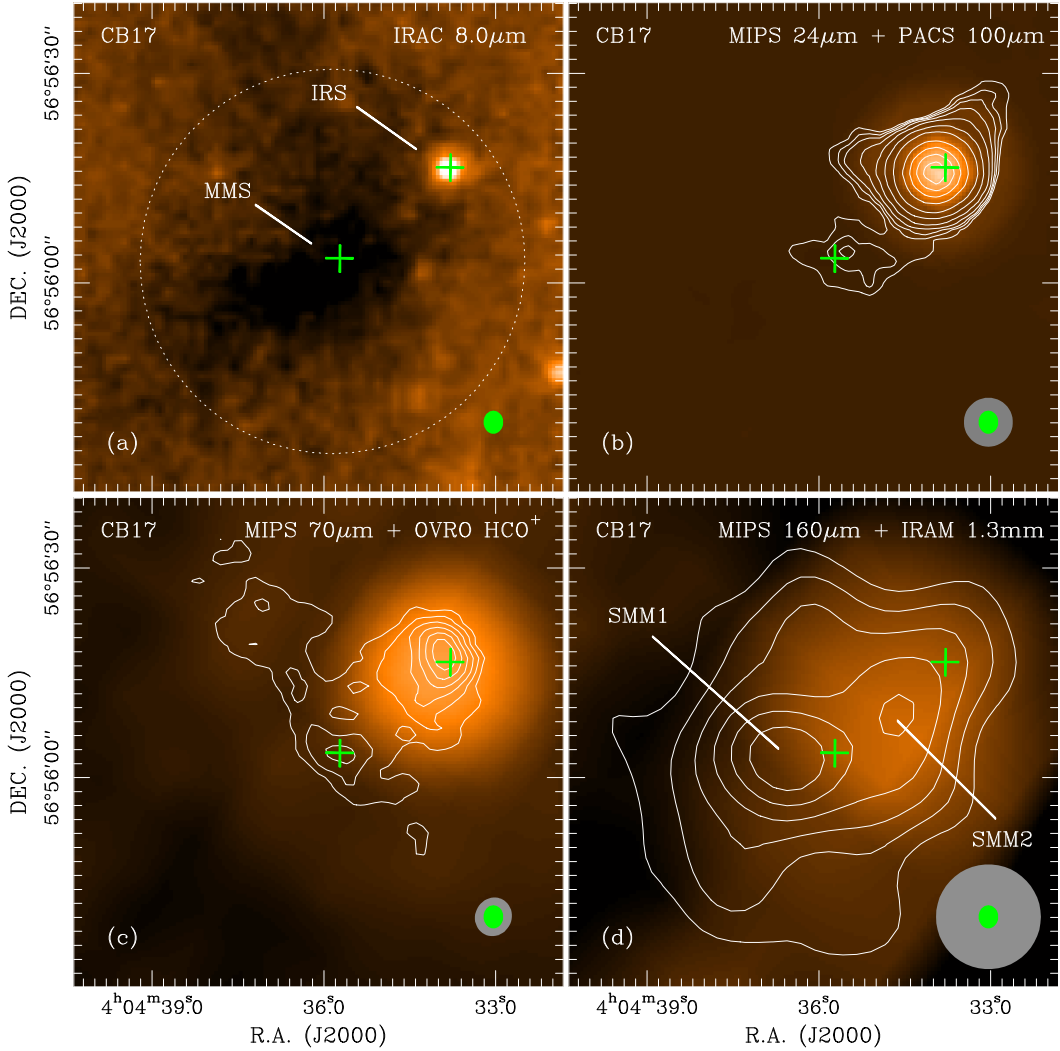}
\caption{$Spitzer$ IRAC and MIPS images of CB\,17. The two green crosses show the positions of the two SMA
dust continuum sources (IRS and MMS), and the synthesized SMA beam is shown as a green oval in the bottom 
right corner. (a) IRAC 8.0\,$\mu$m image. (b) MIPS 24\,$\mu$m image, with the {\it Herschel} PACS 100\,$\mu$m 
continuum contours (white) overlapped (data from M.~Schmalzl et al., in preparation). The PACS contours correspond 
to 100, 105, 110, 120, 140, 180, 250\,mJy\,beam$^{-1}$, and then increase in steps of 150\,mJy\,beam$^{-1}$. The
grey oval in the bottom right corner shows the angular resolution in the PACS 100\,$\mu$m observations ($\sim$\,7$''$). 
(c) MIPS 70\,$\mu$m image, with the OVRO HCO$^{+}$\,(1--0) emission contours (white) overlapped (data from 
M.~Schmalzl et al., in preparation). The HCO$^+$ contours (integrated over the velocity from $-$5.3 to 
$-$4.1\,km\,s$^{-1}$) start at $\sim$\,3\,$\sigma$ and then increase in steps of 1\,$\sigma$ 
($\sim$\,50\,mJy\,beam$^{-1}$\,km\,s$^{-1}$). The grey oval in the bottom right corner shows the synthesized OVRO 
beam size ($\sim$\,5$''$). (d) MIPS 160\,$\mu$m image, with the IRAM-30m 1.3\,mm dust continuum contours (white) 
overlapped (data from Launhardt et al. 2010; smoothed to 15$''$); the IRAM 1.3\,mm contour values are 20, 26, 32, 
38, 44,  and 50\,mJy\,beam$^{-1}$. The grey oval in the bottom right corner shows the smoothed IRAM-30m beam size.
\label{cb17_spitzer}}
\end{center}
\end{figure*}


\begin{figure*}[hlpt]
\begin{center}
\includegraphics[width=15cm,angle=0]{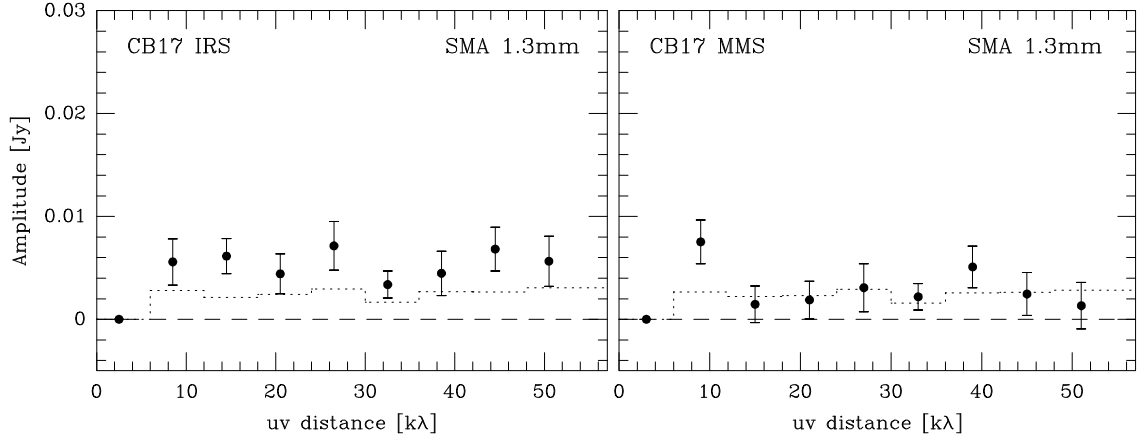}
\caption{Visibility amplitude vs. $uv$ distance plots for the continuum sources CB17\,IRS ($Left$) 
and MMS ($Right$) with 1\,$\sigma$ error bars. The dotted histograms are the zero expectation 
level.\label{cb17_uvamp}}
\end{center}
\end{figure*}

\begin{figure*}
\begin{center}
\includegraphics[width=16cm, angle=0]{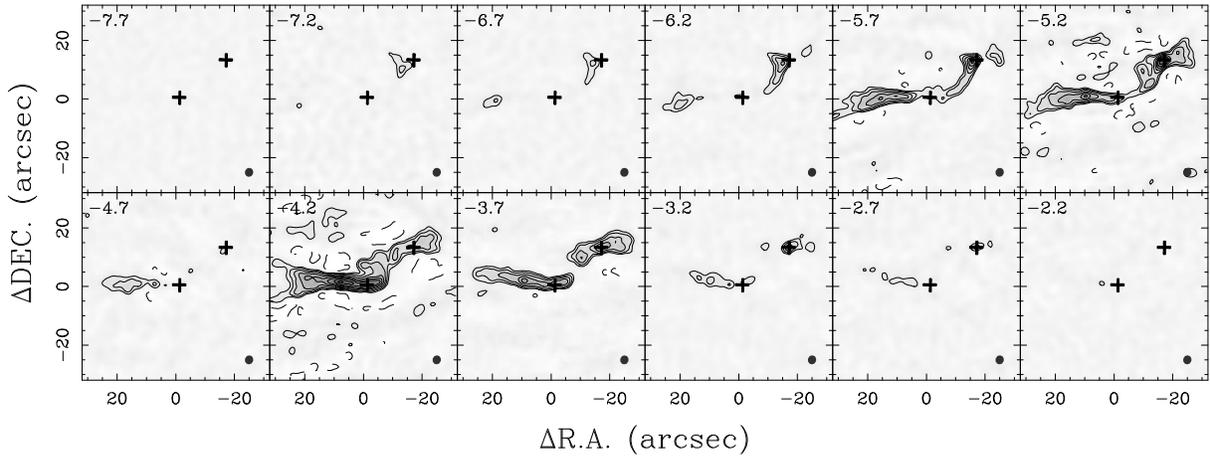}
\caption{Velocity channel maps of the SMA $^{12}$CO\,(2--1) emission of CB\,17 
(phase center R.A.\,=\,04:04:35.85, DEC\,=\,56:56:03.09, J2000). The center velocity 
of the channel is written in top left corner of each panel (in km\,s$^{-1}$). The
systemic velocity of the molecular cloud is $\sim$\,$-$4.7\,km\,s$^{-1}$. Contours levels 
correspond to $-$3, 3, 5, 8, 12, 16, and 20\,$\sigma$, then increase in steps of
10\,$\sigma$, where the 1\,$\sigma$ level is $\sim$\,0.1\,Jy\,beam$^{-1}$. In each panel, 
the two crosses mark the positions of the two SMA dust continuum sources (CB\,17\,IRS 
and MMS), while the filled ellipse (lower right corner) indicates the synthesized beam.\label{cb17_12co21_channel}}
\end{center}
\end{figure*}


\begin{figure*}
\begin{center}
\includegraphics[width=16cm, angle=0]{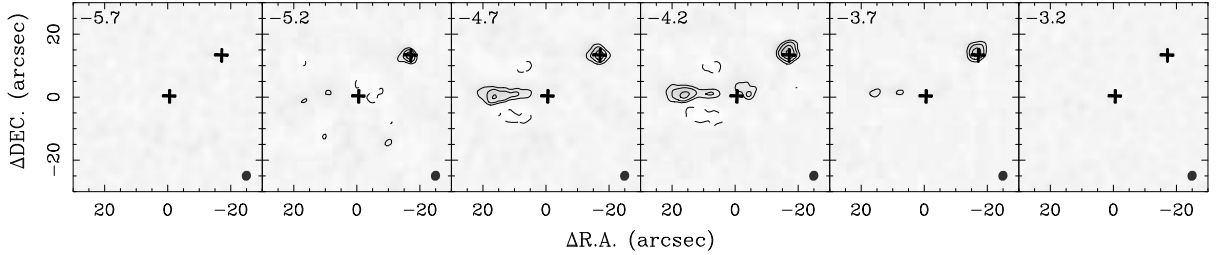}
\caption{Velocity channel maps of the SMA $^{13}$CO\,(2--1)
emission of CB\,17. Contours levels correspond to $-$3, 3, 5, 8,
12, 16, and 20\,$\sigma$, where the 1\,$\sigma$ level is
$\sim$\,0.07\,Jy\,beam$^{-1}$. In each panel, the crosses mark the
positions of the two sources (CB\,17\,IRS and MMS), while the
filled ellipse (lower right corner) indicates the synthesized beam.\label{cb17_13co21_channel}}
\end{center}
\end{figure*}


\begin{figure*}
\begin{center}
\includegraphics[width=5.4cm, angle=-90]{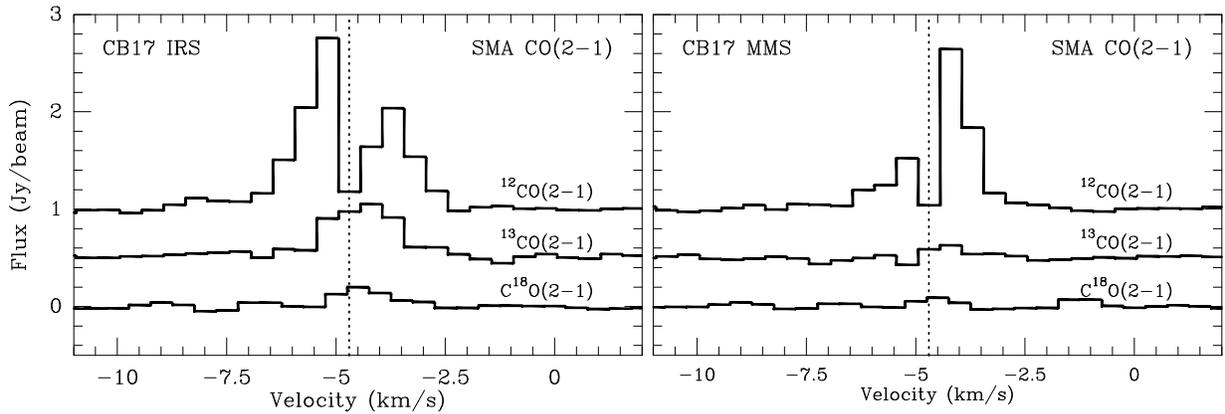}
\caption{SMA CO\,(2--1) spectra at the peak positions of the continuum sources CB\,17\,IRS ($left$) 
and CB\,17\,MMS ($right$). The SMA spectra shown are obtained by averaging the emission within 
3$''$ of the peak position (roughly the beam size). The dotted line indicates the cloud systematic 
velocity derived from the IRAM-30m N$_2$H$^+$\,(1--0) observations ($\sim$\,4.7\,km\,s$^{-1}$).
\label{cb17_smaspectra}}
\end{center}
\end{figure*}


\begin{figure*}
\begin{center}
\includegraphics[width=16cm, angle=0]{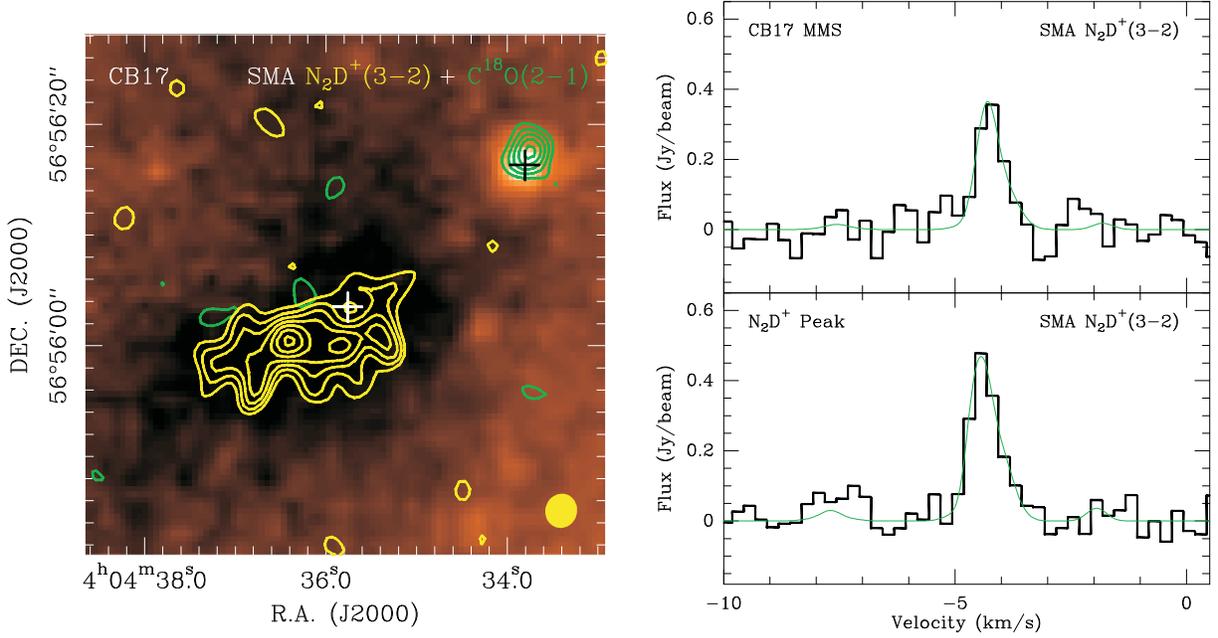}
\caption{{\it Left:} Velocity-integrated intensity maps of the SMA N$_2$D$^+$\,(3--2) emission (yellow 
contours) and C$^{18}$O\,(2--1) emission (green contours), plotted on the IRAC 8.0\,$\mu$m image 
of CB\,17. The N$_2$D$^+$ and C$^{18}$O emission is integrated over the velocity from $-$4.9 to 
$-$3.7\,km\,s$^{-1}$, and contours start at 3\,$\sigma$ and increase in steps of 1\,$\sigma$ 
($\sim$\,0.055 and $\sim$\,0.070\,Jy\,beam$^{-1}$\,km\,s$^{-1}$, respectively). The crosses mark 
the positions of the two continuum sources (CB\,17\,IRS and MMS), while the yellow ellipse indicates the 
synthesized beam in the N$_2$D$^+$\,(3--2) line (3.1$''$\,$\times$\,2.8$''$). {\it Right:} The N$_2$D$^+$\,(3--2)
spectrum (histogram) at the position of source CB\,17\,MMS (top) and at the position of the peak integrated
intensity (bottom), with a model fit of the N$_2$D$^+$\,(3--2) hyperfine structure (continuous line).\label{cb17_n2dp32_map}}
\end{center}
\end{figure*}


\begin{figure*}
\begin{center}
\includegraphics[width=10cm, angle=0]{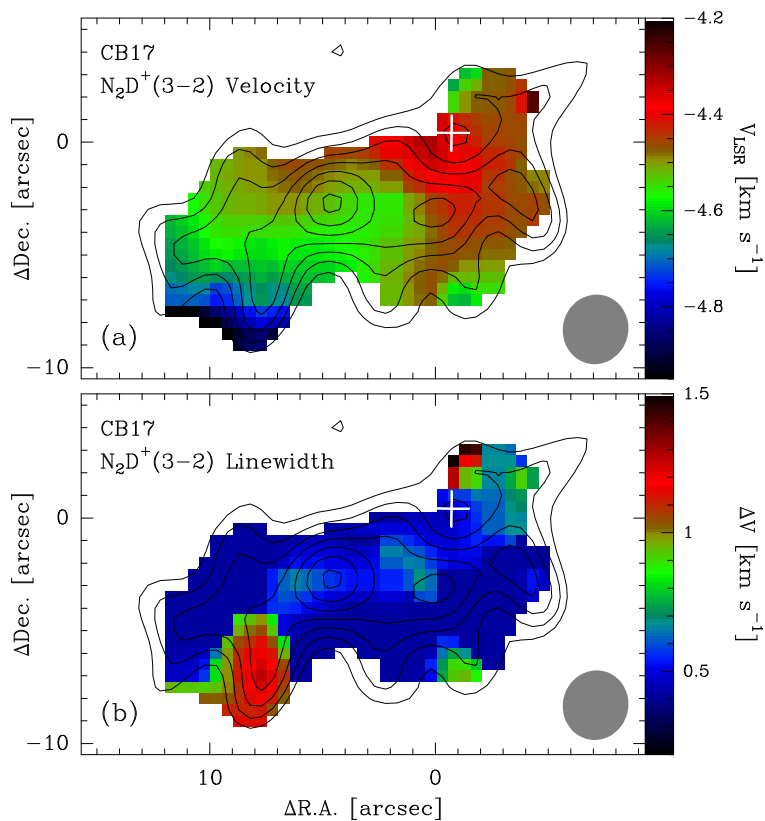}
\caption{N$_2$D$^+$\,(3--2) velocity fields of CB\,17. The top map shows the mean velocity field (color shades), 
while the bottom map shows the spatial distribution of the N$_2$D$^+$\,(3--2) line width (color shades). In both 
maps, black contours show the integrated intensity (from Figure~7), and the cross shows the position of source 
CB\,17\,MMS. The SMA synthesized beam in the N$_2$D$^+$ observations is shown in the bottom right corner.
\label{cb17_n2dp32_velfield}}
\end{center}
\end{figure*}


\begin{figure*}
\begin{center}
\includegraphics[width=12cm, angle=0]{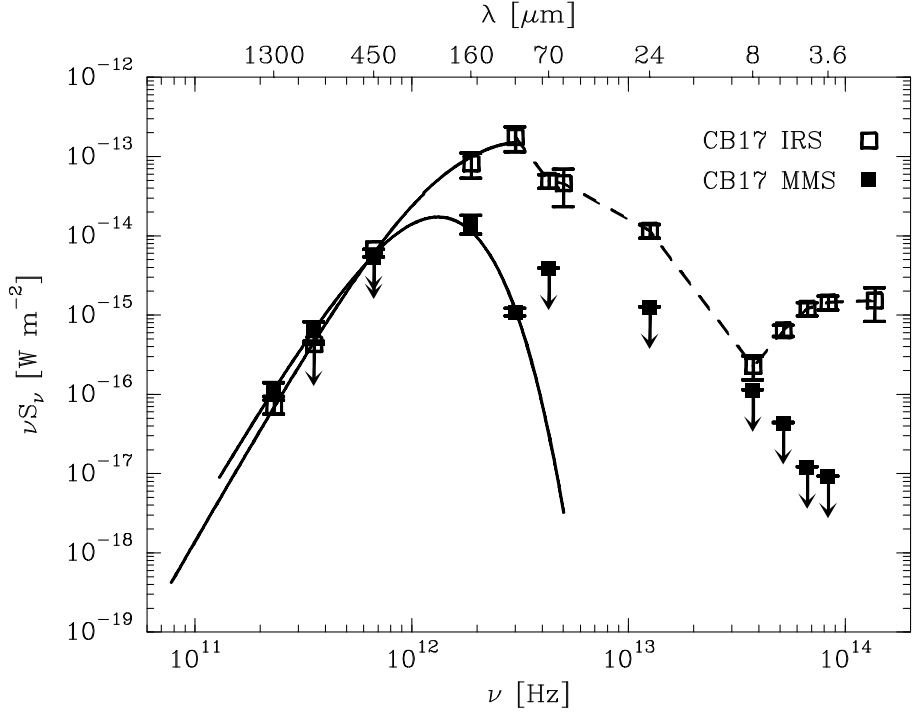}
\caption{Spectral energy distributions of CB\,17\,IRS and MMS (see Table~4 for the values of individual data points).
Solid lines show the best-fit for all points (including upper limits) at $\lambda$ $\geq$ 100\,$\mu$m using a grey-body 
model $S_{\nu}$\,=\,$B_{\nu}$($T_{\rm d}$)(1\,$-$\,$e$$^{-\tau_\nu}$)$\Omega$, where $B_{\nu}$($T_{\rm d}$) is 
the Planck function at frequency $\nu$ and dust temperature $T_{\rm d}$, $\tau_\nu$ is the dust optical depth as a 
function of frequency $\tau$\,$\propto$\,$\nu$$^{1.8}$, and $\Omega$ is the solid angle of the source. Dashed line
shows a simple logarithmic interpolation performed between all points at $\lambda$ $<$ 100\,$\mu$m for source 
CB\,17\,IRS (not performed for CB\,17\,MMS).\label{sed_observations}}
\end{center}
\end{figure*}

\begin{figure*}
\begin{center}
\includegraphics[width=10cm, angle=0]{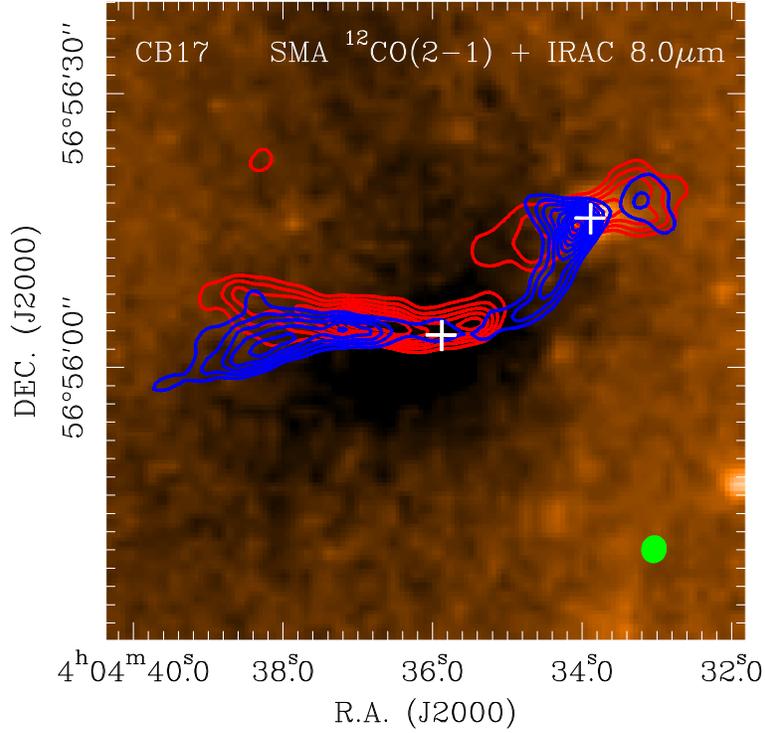}
\caption{The integrated intensity maps of the SMA $^{12}$CO\,(2--1) emission from CB17, plotted on
the $Spitzer$ IRAC 8.0\,$\mu$m image. The blue (red) CO contours represent emission integrated over the 
velocity range $-$7.2\,km\,s$^{-1}$\,$<$\,$V_{\rm LSR}$\,$<$\,$-$5.7\,km\,s$^{-1}$
($-$3.7\,km\,s$^{-1}$\,$<$\,$V_{\rm LSR}$\,$<$\,$-$2.2\,km\,s$^{-1}$), which is blueshifted
(redshifted) with respect to the cloud systemic velocity ($-$4.7\,km\,s$^{-1}$). The CO contours start
at $\sim$\,0.45\,Jy\,beam$^{-1}$\,km\,s$^{-1}$ ($\sim$\,3\,$\sigma$), and increase in steps of 
0.3\,Jy\,beam$^{-1}$\,km\,s$^{-1}$ ($\sim$\,2\,$\sigma$). The two white crosses mark the positions of the
two continuum sources (CB\,17\,MMS and IRS), while the green oval in the bottom right corner indicates the 
SMA synthesized beam ($^{12}$CO line).\label{outflow}}
\end{center}
\end{figure*}

\begin{figure*}
\begin{center}
\includegraphics[width=10cm, angle=0]{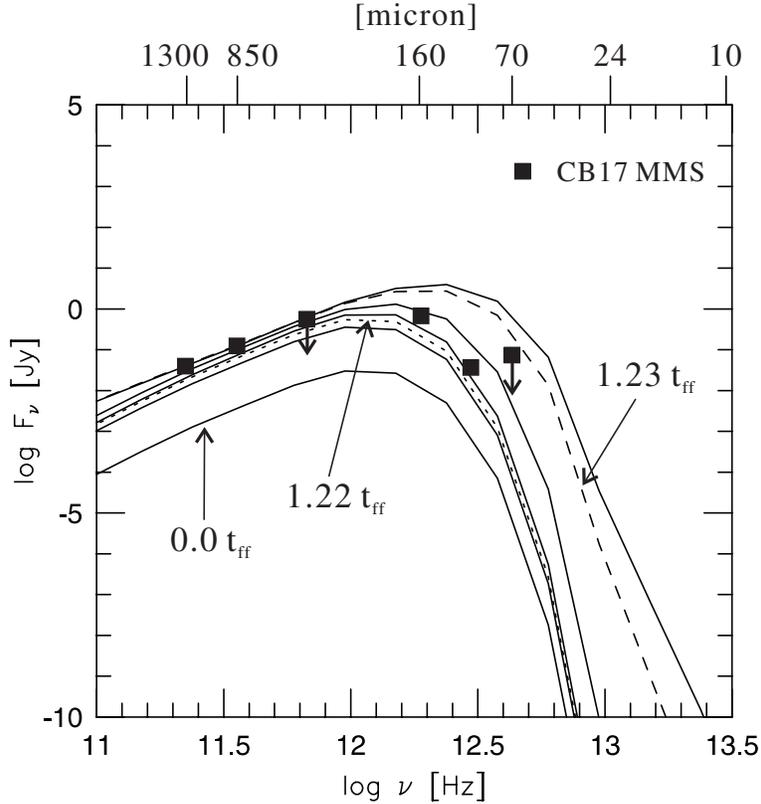}
\caption{Comparison between the fluxes of CB\,17\,MMS (see Table~4) and the fluxes expected for a first core at 
different evolutionary stages (fluxes that would be observed with the beam size of 10$^{3}$\,AU or 4$''$ at the 
distance of CB\,17; from the model M1a in Masunaga et al. 1998). The solid, dashed, and dotted lines show the 
evolution of the SED of a first hydrostatic core with a luminosity of 0.06\,$L_\odot$ (evolved at $\rho$$_{\rm center}$ 
$\sim$ 10$^{-9}$\,g\,cm$^{-3}$ or $time$ approximately 1.23\,t$_{\rm ff}$), where the t$_{\rm ff}$ is the free-fall time. 
Note that the beam sizes of SCUBA and IRAM-30m (see Table~4) are larger than the beam size used to derive 
fluxes in the model.\label{sed_model}}
\end{center}
\end{figure*}

\end{document}